\documentstyle[12pt]{article}

\begin{document}
\begin{flushright}
Preprint CNS-9707\\
April 1997\\
\end{flushright}

\begin{center}
\LARGE
{\bf Statistical analysis of scars in stadium billiard}
\vspace{6mm}\\
\small
Baowen Li$^{a,b}$\footnote{E-mail: 
baowenli@hkbu.edu.hk} and Bambi Hu $^{a,c}$\vspace{6mm}\\
$^{a}$Department of Physics and Centre for Nonlinear Studies, Hong Kong 
Baptist University, China \\
$^{b}$ Center for Applied Mathematics and Theoretical Physics,
University of Maribor, Krekova 2, 2000 Maribor, Slovenia\\
$^{c}$ Department of Physics, University of Houston, Houston TX77204, USA
\vspace{10mm}\\
\end{center}

\begin{center}
\large{\bf Abstract}
\end{center}
\bigskip
\bigskip

\normalsize
\noindent
In this paper, by using our improved plane wave decomposition method, we 
study the scars in the eigenfunctions of the stadium 
billiard from very low state to as high as about the one millionth state.
In the
systematic searching for scars of various types, we have used the approximate
criterion based on the quantization of the classical action along the unstable
periodic orbit supporting the scar.
We have analized the profile of the
integrated probability density along the orbit. We found that the
maximal integrated intensity of different types of scars scales 
in different way with the $\hbar$, which confirms 
qualitatively and quantitatively the existing theories of scars such as 
that of Bogomolny (1988) and that of Robnik (1989). 
\vspace{10mm}
\\\\
PACS numbers: 05.45.+b, 03.65.Ge, 03.65.Sq
\bigskip
\\\\
To appear in {\bf Journal of Physics A} {\bf 31} 1998 
\newpage

\section{Introduction}

In the study of quantum chaos, energ level statistics and
wavefunction statistical properties are of great 
fundamental importance. They are 
proper measures to describe the signature of chaos in a quantum system 
whose classical counterpart is chaotic. After unfolding, the energy level 
statistics has some universal behaviours in the semiclassical limit. It 
has been conjectured by Bohigas {\it et al}\cite{BGS84} that the level 
fluctuations only depend on general space-time symmetry, and are the same 
as predicted by the Random Matrix Theory. Extensive numerical and 
experimental results have supported this conjecture (see e.g. 
\cite{Bohigas89}), although a rigorous mathematical proof of this 
conjecture is still missing.

However, in spite of the importance, the wavefunction of a quantum chaotic 
system has so far remained a relatively less studied area as compared to 
the energy spectra. A counterpart of WKB-Ansatz, which valids in the case 
of an integrable system, is still missing for a chaotic system.   
The only proven 
result is the so called "Shnirelman's theorem" \cite{Sh74}, which 
deals with the
phase-space measures associated to eigenstates of a classically ergodic
system in the semiclassical limit. Shnirelman's theorem predicts that as 
the 
energy goes to infinity, the probability density of most eigenstates of a 
chaotic billiard approaches an uniform distribution.
This is consistent with the  prediction of Berry \cite{Berry77} and 
Voros\cite{Voros79}. One major surprise is the discovery of the strong 
enhancement of the probability density along the least unstable periodic 
orbits, which was first observed by McDonald and Kaufman\cite{McDkauf}, 
and 
later also by Heller\cite{Heller84} for stadium billiard. This 
kind of structure was named "scar" by Heller.

Since its discovery, much efforts has been contributed 
to understand this interesting phenomenon in the last decade, and much
progress has been achieved up to now.
On the theoretical side, Bogomolny \cite{Bog88} 
developed a semiclassical theory of scars in configuration space, and
Berry \cite{Berry89} performed a similar analysis in phase space
using the Wigner function. According to this theory, the intensity (see 
Eq.(\ref{int}) for the definition) of a scar goes as $\sqrt{\hbar}$. 
Furthermore, based 
on the semiclassical evaluation of the Green function of the Schr\"odinger
equation in terms of the classical orbit, Robnik \cite{Robnik89} has 
developed a theory, which suggests that although the geometrical 
structure of the scar can be determined by a single short periodic orbit 
(primary orbit), 
the maximal intensity of the scar is nevertheless determined by the sum 
of contributions from similar but longer periodic orbits, which 'live' 
in the homoclinic neighbourhood close to the stable and unstable 
manifolds of the primary orbit. And the maximal intensity is independent of
$\hbar$. 
The contribution of homoclinic orbits surrounding the primary orbit to 
the density of states 
has been studied by Ozorio de Almeida\cite{Almeida89}.
Most recently, Klakow and Smilansky 
\cite{KLSM} 
used a scattering quantization approach to study the scar problem. 
Parallel to 
the theoretical developments, there have also been many numerical
\cite{WYMGR,AGFI,ProRob} and experimental\cite{SRHE} studies.

Unfortunately, due to the limit of the numerical techniques and the
computer facilities, most of the numerical studies so far are
undertaken only at very low energy range, which is too low to verify 
the theoretical predictions in the very far semiclassical limit, 
especially for Robnik's theory. 

In this paper, we propose a new numerical technique for solving the 
eigenvalue problem of 2-D stadium billiard. Since our method is 
based on 
the Heller's plane wave decomposition method (PWDM), we call it 
the improved PWDM. (For more
details about Heller's PWDM, please see \cite{Heller89} and \cite{LR}.) 
By using this improved PWDM, we have been  
successful to go to as high as the 1 millionth state, which we believe is 
already very deep in the semiclassical regime for the stadium billiard.
Moreover, with the help of the semiclassical criterion \cite{Bog88,Robnik89},
we have found many consecutive scars in several different energy ranges,
which spans 2 orders of magnitude in the wave number. With this collected 
ensemble of scars, we are able to study many properties of scars such as 
the scaling 
property of the scar intensity  profiles with $\hbar$ up to the very far 
semiclassical limit. 

The paper is organized as follows. In Sec. 2 we discuss the improved PWDM
which is used in calculating all the high-lying eigenstates discussed in this
paper. The properties of different type of scars are discussed in Sec. 3.
In Sec. 3.1, we discuss the scar type, whose
maximal integrated intensity is independent of $\hbar$, which 
evidently supports Robnik's
theory of scars; while in Sec. 3.2 and 3.3  we discuss the scar type, its
geometrical structure can be predicted by Bogomolny's theory very well. More
examples of scars and the bouncing ball states are also briefly discussed in 
Sec. 4. We end our paper by discussions and concluding remarks in Sec. 5. 
Part of the works in Sec. 3.1 was reported earlier \cite{Li97}. 

\section{The improved plane wave decomposition method}

As was mentioned previously, the difficulty of studying the 
eigenfunctions in 
the very far semiclassical limit lies in the numerical calculation of the 
eigenenergies and the corresponding eigenfunctions. The usual 
diagonalization method is 
not suitable because it calculates all the eigenvalues from the 
ground state up to a certain eigenenergy. Therefore, the dimension of 
the matrix to be 
diagonalized increases with the sequential number of the 
eigenstates.  This drawback becomes the greatest obstacle
if we want to go to the regime very far in the semiclassical limit. 
Among many other methods, Heller's plane wave decomposition method is most 
suitable one for the study of the high-lying eigenstates. In previous works, 
Li and Robnik \cite{LR} have 
used this method to calculate the eigenstates as high as the 200,000th 
states in a KAM and a chaotic billiard. 
However, in order to go to even higher energy, this technique runs into 
difficulty of spending too much CPU time on the matrix inversion. Thus, 
it is necessary to improve this method to allow us to go much higher in the 
semiclassical limit with nowadays suitable computer facilities. 
As we shall see in the following that our improved PWDM is at least 
5 times faster than the PWDM, which
make it possible to test the semiclassical theory of scars. 

To solve the Schr\"odinger equation with Dirichlet boundary conditions

\begin{equation}
\Delta \Psi + k^2 \Psi =0,\quad \Psi=0 \quad \mbox{at the boundary}
\label{Schreq}
\end{equation}

\noindent 
we use the superposition of plane waves with the wave vectors of the same 
magnitude $k$ but with different directions. The wavefunction we used for 
the odd-odd parity of the stadium billiard is

\begin{equation}
\Psi(x,y) = \sum_{j=1}^{N} a_j\sin(k_{jx}x)\sin(k_{jy}y),
\label{WVFunc}
\end{equation}

\noindent
where $k_{jx} = k\cos(\theta_j), k_{jy}=k\sin(\theta_j)$, $k^2 =E$ is the 
eigenenergy, $N$ the number of plane waves, $\theta_j= 2j\pi/N$, i.e. the 
direction angles of the wave vectors are chosen equidistantly. The {\it 
ansatz} 
(\ref{WVFunc}) solves the Schr\"odinger equation (\ref{Schreq})  inside 
the billiard, so that we have only to satisfy the Dirichlet boundary 
condition.  For a given $k$, we set the wavefunction equal to zero at a 
finite number $M$ of boundary points (primary nodes) and equal to 1 at an 
arbitrary chosen interior point. It is obvious that in order to avoid the 
underdetermined problem we should take $M \ge N$. This gives an 
inhomogeneous 
set of equations which can be solved by matrix inversion. Usually the 
matrix is very singular, thus the singular value decomposition method has 
to be invoked. After obtaining the coefficients $a_j$, we calculate the 
wavefunctions at other boundary points (secondary nodes). The sum of the 
squares of the wavefunction at all the secondary nodes (Heller called 
it "Tension")  would ideally be zero if $k^2$ is an eigenvalue. In 
practice, it is a 
positive number. Therefore, the eigenvalue problem becomes to be finding the 
minimum of the "tension". In practical implementation, it is better to 
look for the zeros of the first derivative of the tension (for 
convenience we denote this function with $f(k)$), since 
the derivative is available analytically/explicitly from (\ref{WVFunc}) once 
the coefficients $a_j$  have been found.

This is the main idea of Heller's method. In general, this method takes
several (usually about 10, it depends on the stepsize) iterations to 
find an eigenvalue, which means that about 10 matrix inversions must be 
performed. This costs a lot of CPU time and turns to be the main 
shortage of this method. So, the primary motivation of 
our new technique is to reduce the 
number of 
the matrix inversions. As we shall see soon, this can be achieved 
without any difficulty.

Since we have already calculated the coefficients $a_j$ 
after one matrix inversion,
the function $f(k)$ can be expanded into Taylor series around $k_0$
\begin{equation}
f(k) = f(k_0) + \sum_{n=1} \frac{f^{n}(k_0)}{n!}(k-k_0)^n ,
\label{Ften}
\end{equation}
where $f^n(k_0)$  is the nth derivative of $f(k)$ at $k_0$, which can be 
calculated analytically/explicitly very easily. Thus, our 
task now is to find the roots of this polynomial, which, as it is well 
known,  
costs much less CPU time than the matrix inversion. 
Then, the eigenvalue 
around $k_0$ is approximately equal to $k_0 + \Delta k$, where $\Delta k$ 
is the smallest root of the polynomial (\ref{Ften}). Our numerical 
experience 
demonstrate that with this improved method, we can get the eigenvalue 
with accuracy of less than one percent of mean level spacing by just 
doing one matrix inversion. To get higher accuracy, we should use the 
new eigenvalue $k$ and do further matrix inversion. Then calculate 
the new coefficients $a_j$, and find out the smallest root of the 
new polynomial. This procedure can be continued until an expected 
accuracy is reached. In our numerical calculations, 
for almost all the cases, by doing about 2-3 matrix inversions we may get 
the eigenvalue with accuracy as high as $10^{-4}$ of the mean 
level spacing. Therefore, our improved PWDM reduces the CPU time about 5 
times or more as compared to the original Heller's PWDM. In our 
practical implementation,
 the function $f(k)$ is expanded up to the 8th to the 10th order, which is 
already good enough to obtain above mentioned accuracy.

Before proceeding to do any analysis of the scars, let us spend a few 
words to discuss how to search and collect the scarred eigenstates 
systematically and extensively, because we need enough ensembles of the 
scarred eigenstates to make the numerical analysis  significant.
Therefore, our first step is to collect scars of the same type in a 
wide range of energy. We 
begin from a very low state, e.g. from the ground state. As long as we 
find the first scarred state, say, e.g. at the wave number $k_0$, then 
we can use the semiclassical criterion to estimate the next scar.
According to the semiclassical theory \cite{Bog88,Berry89,Robnik89}, 
the scar will most likely occur if quantized, i.e
\begin{equation}
S =2\pi\hbar\left(n+\frac{\alpha}{2}\right),\qquad n=0,1,2,... ,
\label{action}
\end{equation}
$S$ is the action along the periodic orbit, $\alpha$ the Maslov phase.
Thus, we jump to the wave numer at about $k= k_0+\Delta k$ to calculate
the eigenvalue and eigenfunction, where $\Delta k = 2\pi\hbar/{\cal L}$,
${\cal L}$ is the length of the periodic orbit. Usually, we need to
calculate a few eigenstates around $k$ to locate the scarred eigenstate. We
continue this procedure until we collect a satisfied ensemble of scars. 
It is shown that, this procedure is very helpful in estimating the energy
range of the scarred state at the very far semiclassical limit. For
instance from a very low scarred eigenstate at $k_0$ we can skip over a
very large number of states to a rather high level, e.g at $k= k_0 +
m\Delta k$. $m$ may be a very large number e.g. about a few hundred.  As we
shall see later that in many cases this criterion is even accurate  
within
one mean level spacing, namely, the scar occurs at the eigenstate whose
eigenenergy is roughly equal to the predicted energy by this way. 

However, it must be pointed out that, the semiclassical theory
Eq.(\ref{action}) cannot predict the individual state at which the scar
will occur. Instead, as mentioned before, if we have already found one
scar, say at $k_0$, then the semiclassical theory just tells us that the
eigenstates at the wave number of $k_0 \pm \Delta k$, will most likely be
scarred. 

In our study we put $\hbar=1$, so, the inverse of wave number $k$ plays
the role of $\hbar$, i.e. $k$ goes to infinity indicates the semiclassical
limit. 

\section{Statistical analysis of scars}

We would like to do quantitative analysis of the scars in this section. As
already mentioned in previous section, in order to do any significant
statistics we should have enough ensemble of scars of the same type. In
searching and collecting the scarred eigenstates we use both qualitative
and quantitative procedures. We start from very low state and calculate
the probability density plots of wavefunction for many consecutive
eigenstates, usually in the order of 20. We judge at first by eyes whether
the states is scarred or not by a certain kind of unstable periodic orbit
(PO), e.g. diamond shape PO or the horizontal PO. Generally, this
procedure is quite accurate and reliable, although it is qualitative.
Furthermore, in order to improve the objectiveness of the judgment, we
calculate the integral intensity according to Eq.(\ref{int}) to check
which scarred eigenstate is the most favourite candidate. Relying on
these two procedures we are able to select our scarred eigenstate very
objective and with high reliability. As long as the first scarred
eigenstate is determined, we may use the semiclassical criterion
Eq.(\ref{action}) to chose the energy range in which the next scarred
eigenstate will most likely occur. Then repeat the procedure mentioned
above and find out the next scarred eigenstate. In this way, we were able to
select a sufficient number of scarred eigenstates from a huge number
eigenstates (about 10,000 eigenstates ranging from very low to about 1
millionth state) for our numerical analysis. The quantitative analysis is
given in the following. 

\subsection{Scars supported by the diamond shape periodic orbit}

In this section, we shall discuss a type of scar which demonstrates that
the maximal integrated intensity never vanishes as $\hbar$ goes to zero.
This finding is very different from the prediction of the common believed
theory- single periodic orbit theory, but it can be explained by Robnik's
theory, as we shall see later. The main results of this part have been
reported earlier in Ref.\cite{Li97}, but more details about the 
wavefunction structures are given here. 

With the help of the semiclassical quantization criterion Eq.
(\ref{action}) and the procedure described above, we have gathered about 100
examples of the same type scarred eigenstates at different energy ranges,
namely, $k$ ranges from about $10$ to $k\approx 1330$. Here, we select
only 6 representatives of these scarred eigenstates from the very low
states to the very high states. They are shown in Fig. 1a-f. The eigen
wave numbers are given at the top of each figure. The lowest one,
$k=$10.240 95 corresponds to about the 40th eigenstate, while the highest
one $k=$ 1328.153 849, corresponds to the sequential number 250,034 for
odd-odd parity, and to the index about 1,001,408 when all parities are
taken into account. To the best of our knowledge, this is the highest
eigenstate showing significant scar so far. 

Suprise as it is, in addition to the eigenstate shown in Fig. 1f, we have
also collected quite a few examples of this type of scarred states in such a
high energy. This implies that this type of scar survives the
semiclassical limit. One may ask: does this finding contradict with the
Shnirlman's theorem\cite{Sh74}, which states that as the energy goes to
infinity, the probability density of most eigenstates of a chaotic
billiard approaches a uniform distribution? To test this, we have examined
the statistics of the probability distribution function of the eigenstate,
and found that it is an excellent Gaussian distribution function, although 
there is such a pronounced density around the unstable periodic orbit. The
probability distribution function $P(\Psi)$ ($P(\Psi) d\Psi$ is the
probability of finding the wavefunction of value $\Psi$) as well as the
cumulative distribution function $I(\Psi) = \int_{-\infty}^{\Psi} P(t)dt$
are shown in Fig. 2a and 2b, respectively.  They are compared with the
theoretical values of the Gaussian random model\cite{Berry77} which predicts
\begin{equation}
P(\Psi) 
=\frac{1}{\sqrt{2\pi}\sigma}\exp\left(-\frac{\Psi^2}{2\sigma^2}\right).
\label{Gaussian}
\end{equation}
Even if we magnify the small details in the cumulative figure as shown in
the boxes of Fig. 2b, the discrepency with the Gaussian function is almost
indistinguishable. Where $\sigma^2=1/{\cal A}$, should be equal to
$\langle\Psi^2({\bf x})\rangle$, the average probability density inside
billiard, according to the semiclassical theory
\cite{Sh74,Berry77,Voros79}. ${\cal A}$ is the area of the billiard. 

In order to understand the scar properties quantitatively, we have
investigated the following pronounced (excess) intensity in a thin tube
along the periodic orbit (see Fig. 3), which is defined by
\begin{equation}
I = \frac{\int \Psi^2({\bf x}) 
d{\bf x}}{\int \langle\Psi^2({\bf x})\rangle d{\bf x}} - 1 ,
\label{int}
\end{equation}
where $\Psi({\bf x})$ is the eigenfunction at ${\bf x}$.  The integral is
taken over a thin tube around the periodic orbit as is shown in Fig. 3. 

In Fig. 4a-f, we display the integrated intensity (\ref{int}) versus the
width of the tube ($D$) in unit of the de Broglie wavelength around the
periodic orbit for the scarred states shown in Fig.1. The wave number of
each state is given at the left bottom of the box. 

The first thing to be seen from these profile figures is that the scar
intensity reaches a maximum at the width of about the 1--2 de Brodglie
wavelengths from the periodic orbit. This agrees with Robnik's theory
which states that the semiclassical waves associated with individual
daughter orbits interfere constructively with each other only within a
tube of width 1--2 de Brodglie wave length. The second important feature
of these figures is that the magnitude of the maximum does not change too
much although the eigenenergy changes more than 100 times. 

Furthermore, after checking the eigenenergies of these 6 examples
carefully, we found that the semiclassical criterion works very well as
mentioned in Sec. I, even though we go from one scarred state to another
one by jumping even up to a few hundred scarred states.  For instance,
starting from the first eigenvector $k_0$ = 10.241 095, if we go through
65 scarred states, we have $k = k_0 + 65\Delta k =$ 101.563 684, this
value is very close to the true eigenvalue $k_{exact} =$101.568 640.
(Please note that, in this paper, we study only the eigenstates with
odd-odd parity, so the length of the periodic orbit shown in Fig. 3 is
${\cal L} = 2\sqrt{5}$ rather than $4\sqrt{5}$ for the total billiard,
thus, $\Delta k = 2\pi/{\cal L} =$ 1.404 96.) The deviation is less than
one mean level spacing. This procedure applies also to many other scarred
states and it can be verified readily for other states given in Fig. 4.
The validity of the semiclassical criterion for the scarred eigenstates
discussed here has also been verified very recently by Frischat and
Doron\cite{FriDor97} in studying the scars occuring in a quantum system
having a mixed classical dynamics, where regular and irregular region
coexist in the classical phase space. 

Now we turn to an important question, namely, the energy or $\hbar$
dependence of the maximal integrated intensity. This is a rather difficult
problem, even in numerical calculations. Our numerical results show that
around a certain $k$, the maximal integrated intensity varies from the
scarred state to state. This property is clearly shown in Fig. 5, where we
plot 26 consecutive scarred states around $k=125$, all of these 26
eigenstates show very significant localization of wavefunction around the
periodic orbit. One interesting thing should be noted from this plot is that,
there are two cases, one at $k\approx 121$ and the other at $k\approx 125$
showing that two consecutive eigenstates are nearly degenerate, thus both
of them are scarred. Again, from this figure we can also see clearly that
the semiclassical criterion(\ref{action}) works excellently, namely,
the interval of the wave number between two scarred states is almost a
constant, which is approximately equal to $2\pi/{\cal L}$.  The maximal
integrated intensity, however, fluctuates from state to state, which can
not be explained by any existing semiclassical approaches.  This is still
an open problem deserve further theoretical and numerical investigations. 

The results given in Fig. 5 implies that in order to make the study of
dependence of the maximal integrated intensity on energy significant, we
should take certain kinds of ensemble averaging. In our numerical study,
we have performed such averaging around a certain $k$ over many scarred
eigenstates (usually about 10 scarred states).  The averaged results are
drawn in Fig. 6. The least-square fitting gives rise to
\begin{equation}
\langle I_{m}\rangle = 0.73/k^{\alpha},\qquad \alpha = 0.06\pm 0.03,
\label{Iscar1}
\end{equation}
where $\langle.\rangle$ is the local average over many scarred states.
Obviously, the exponent $\alpha=0.06$, which is very close to zero, is far
from $1/2$ predicted by Bogomolny's theory. This fact indicates that the
maximal integrated intensity does not depend on the energy or the $\hbar$
for the scar type shown and discussed in this section. This discovery is
very different from previous one \cite{AGFI} and cannot be explained by
the semiclassical theory of Bogomolny \cite{Bog88} and Berry
\cite{Berry89}, however, it confirms quantitatively the theoretical
prediction of Robnik \cite{Robnik89}, which states that the maximal
intensity of a scar, is independent of $\hbar$, if the scar is supported
by many orbits as mentioned above. 

There are two important ingredients in Robnik's unpublished theory: (1)
The width of the scar profile is about the order of the de Broglie
wavelength; (2) There are many similar longer periodic orbits contributes
to the scar intensity. The first one comes from a very simple physical
argument. The scar profile cannot be smaller than the de Broglie
wavelength since this is the smallest scale at which the quantum waves
explore the classical dynamics. However, it can neither be much larger
than that scale, simply because the contribution of the geometrically
similar but longer periodic orbits would destroy the scar beyond the
distance of one de Broglie wavelength, as the waves would interfere
destructively there, while they would interfere constructively within 
the region of order of one de Broglie wavelength.  As to the
second point, the reason is that the periodic orbits, close to the stable
and unstable manifolds and in the vicinity of the primary periodic orbit,
complete at first a few quasi-cycles which are very close to the primary
orbit, and only then diverge away before the final and ultimate closure.
So, the first few approximate cycles of such longer orbits do resemble the
primary periodic orbit, but they do not close exactly. The excursion of
such orbits away from the primary orbit implies for the semiclassical
waves an unavoidable loss of phase coherence beyond the distance of the
order of one de Broglie wavelength away from the maximum of the scar.
Taking into account all these orbits, the pronounced intensity of the scar
defined by Eq.(\ref{int}) can be described by the following formula,
\begin{equation}
I \approx \nu \sum_{n=1}^{\infty} \frac{sin(n S_1/\hbar)}
{sinh(n\lambda\tau/2)} - 1 ,
\label{Robnik}
\end{equation}
where, $S_1$ is the action along the primary periodic orbit, $\lambda$ is
the Lyapunov exponent of the primary orbit with the period of $\tau$,
the summation over $n$ is due to the repetitions of the orbit and $\nu$ is
the number of contribution orbits, which is determined by criteria of
correct phasing.  Eq.(\ref{Robnik}) tells us that {\em the maximal
intensity} of the scar, when supported by many periodic orbits, is
independent of $\hbar$.  Finally, we would like to point out another
important factor of Robnik's theory, i.e. in deriving 
Eq.(\ref{Robnik}), the averages have been taken over only one mean level 
spacing. Therefore, Eq.(\ref{Robnik}) generally applies to the individual 
eigenstates. This is different from the theory of Bogomolny which we shall
discuss later. 

Our numerical results presented in this section provide the
first and very significant evidence supporting Robnik's theory.  In next
section we shall discuss another type of scars which display a very
different behaviour. 

\subsection{Scars supported by the V shape periodic orbit}

The theoretical prediction from Robnik is different from that of
Bogomolny. We should say that, however, it does not contradict with that of 
Bogomolny at all.
Instead, it is an extension of Bogomolny's theory to the scars supporting
by many periodic orbits. These two theories describe different type of
scars.  In fact, there have already been some numerical results supporting
Bogomolny's theoretical prediction \cite{AGFI}, although these numerical
calculations are limited to very low states. 

By employing our improved PWDM, we are able to go to much higher than before
and to test Bogomolny's theory. In our numerical investigation, in 
addition to the scars discussed in
previous section, we have also obtained other type of scars whose maximal
intensity scales with $\hbar$ in a very different way from that one
given in (\ref{Iscar1}) and (\ref{Robnik}). 

Using the same strategy, i.e. making use of the semiclassical quantization 
criterion, we
have collected a few dozens of the scars of the same type. One
representative in the far semiclassical limit is shown in Fig. 7. The scar
is obviously supported by the V-shape unstable periodic orbit. (This type
of scar was also observed by Heller
\cite{Heller84} at the very low state.) The wave number of the eigenstate in Fig. 7 is
$k=$1328.093 482, which corresponds to the index 250,012 (odd-odd), and to
the index about 1,001,310 for the total billiard. Again, to test
Shnirelman's theorem, we have calculated the probability distribution
function. It is shown in Fig. 8. Like that case in Fig. 2, the probability
distribution function is a perfect Gaussian function. The integrated
intensity profile is shown in Fig. 9. The maximal intensity is just about
$0.4$ which is obviously smaller than that of scar type given in Sec. 3.1. 

To look into the $\hbar$ dependence of such type of scars, we 
have made
the local averaging over a few consecutive scarred stated around a certain
wave number $k$, and $k$ changes from about 10 to about 1300. The results
are given in Fig. 10. The least-square fitting result is
\begin{equation} \langle I_{m}\rangle =
1.85/k^{\alpha},\qquad \alpha = 0.24\pm 0.06. 
\label{Iscar2}
\end{equation} 
$\alpha$ differs significantly from zero, thus this type of scars cannot
be described by Robnik's theory. Moreover, it is not difficult to see 
that  this type of scar has some structures. In particular, there exists 
points at which the wavefunction intensity is very high. To understand these 
properties, we shall
invoke Bogomolny's theory. Accordingly, the semiclassical
expression for the wave function is given by\cite{Bog88}: 
\begin{equation}
\langle |\Psi(x',y')|^2\rangle = \rho_0 + \hbar^{1/2} \sum_p 
Im\left[A_p(x')\exp \left(i\frac{S_p}{\hbar} + i 
\frac{W_p^{km}(x')}{2\hbar}y'_k y'_m\right)\right],
\label{Bogpsi}
\end{equation}
the averaging $\langle .\rangle$ is taken over many consecutive
eigenstates (including those unscarred states). For each periodic
trajectory the $x'$ axis is chosen along the trajectory and the $y'_m$
axes are chosen perpendicularly to it.  $S_p=\oint p_n dq_n$ is a
classical action calculated along the trajectory. $A_p(x')$ and
$W_p^{km}(x')$ are classical quantities through the elements of the
monodromy matrix of a given trajectory. (The monodromy matrix of some
shortest periodic orbits are given in Bogomolny's paper \cite{Bog88}.).
Several conclusions can be drawn from this formula: (a) the scar has
finite width perpendicular to the trajectories. It is proportional to
$[\hbar/|W(x')|]^{1/2}$;  (b) the scar strength scales as $\hbar^{1/2}$,
which means that the scar should vanish in the semiclassical limit as
$\hbar \to 0$; (c) there are the so-called self-focal points where the
monodromy matrix element vanishes, i.e. $m_{12}=0$. 

As to the V-shape periodic orbit supporting the scar in Fig. 7, the
self-focal points take place at the position $x'=\sqrt{L(L-R)}$, where $R$
is the radius of the half-circle of the stadium, $L$ the half-length of
the periodic orbit. For the stadium we studied, $R=1$ and for the V-shape
periodic orbit, $L=(1+\sqrt{2})R \approx 2.414$, thus $x'\approx 1.85$.
Here, $x'$ measures the distance from the center of the periodic orbit,
i.e. from the center of the straight line segment of the billiard
boundary.  If we take a look at the wavefunction shown in Fig. 7, we find out
 that there DO exists focal points locating at about
this distance on the periodic orbit. At that point the amplitude of the
probability density of wavefunction is very high.  We believe that this is
a very good example supporting the conclusion (c) of Bogomolny's theory. 
Of course, this is not an accident example coinciding with Bogomolny's
theory. We have more examples exhibiting this structure.

\subsection{Scars supported by the horizontal periodic orbit}

As further evidence, in Fig. 11a-h we present eight examples of scarred
states for stadium with $R=1, \epsilon=0.2$, where $\epsilon$ is the half
length of straight line of the billiard. The scar in these 8 
states is supported by the horizontal unstable periodic orbit. It is very
readily to see that the scar-shape is very similar to that predicted by
Bogomolny\cite{Bog88} (Cf. Fig. 5 of his paper). 

As a quantitative comparison with Bogomolny's theory, we shall first focus
our attention on the position of the self-focal points in these scarred
eigenstates. Roughly the self-focal point situates
at $x'\approx 0.5\sim 0.6$. According to Bogomolny's theory,
the monodromy matrix element $m_{12} = -\frac{2}{R}(L(L-R) -(x')^2)$. In
this case $L=\epsilon + R =1.2$, so that theoretically the self-focal
point should locate at $x'=\sqrt{L(L-R)}=\sqrt{0.24}\approx 0.5$, which is
approximately the case in the wavefunctions shown in Fig 11a-h. 

Furthermore, Bogomolny's theory predicts that the width of the scar 
shrinks with $(\hbar/|W(x')|)^{1/2}$, where $W(x')$  is 
\begin{equation}
W(x') = \frac{2(L-R)}{L(L-R)-(x')^2},
\label{width1}
\end{equation}
for the  horizontal periodic orbit. Since $L=1.2$ and $\hbar 
\sim 1/k$, the width of the scar $D$ is thus proportional to
\begin{equation}
D(x') = \frac{C}{\sqrt{k}}\sqrt{|0.24-x^{'2}|}.
\label{width2}
\end{equation}
Now, we would like to make quantitative comparison by using this formula
(\ref{width2}). In the following calculation, the constant $C$ in Eq.
(\ref{width2}) is determined by ajusting the width of $D$ which is
approximately equals to the scar's width at the lowest scarred state, i.e.
$k=11.994542$. Accidently, the choice of $C = (11.994542)^{1/2}$ gives us
qualitatively the best result. The scar width $D(x')$ for different $k$ is
then calculated by Eq.(\ref{width2}). They are plotted in Fig. 12a-h
corresponding to the eigenvectors $k$ of the eigenstates in Fig. 11a-h. 
Looking at these two set of pictures, we would say that the shape, the
self-focal point and also the width of the scars follow the theoretical
prediction very well. Obviously, the higher the eigenstate, the better the
agreement between Bogomolny's theory and our numerical results. This, of
course, must be the case, because Bogomolny's theory is a semiclassical 
one. 

Having investigated above examples, we are convinced to reach the
following conclusion: the Bogomolny's theory determines not only the
geometry of the scars, but also the intensity profile scaling with
$\hbar$. Finally, it should be pointed out that, strictly speaking,
Bogomolny's theory is based on the averaging over many consecutive states
(see Eq.  (\ref{Bogpsi})), however, our numerical results show that
Bogomolny's function captures the main structure of the individual scarred
eigenstates (see also Heller's lecture \cite{Heller89}). 

\section{Further examples of scars and bouncing ball states}

In addition to the scars illustrated in last section, we have also discovered
quite a lot of scars, supporting by other unstable periodic orbits,
at about the 1 millionth eigenstate. However, because of lacking
sufficient ensembles, we were not able to do the scaling analysis as we
have done in previous section. We just show two examples here. The
corresponding wavenumber are $k=$1328.069 060 and $k=$1328.112 133,
respectively. The sequential number are about 1,000,004 and
1,000,080, respectively, for the total billiard. Evidently, the scar
strength is weaker than that one shown in Section 3.1. It seems
that these scars will not be able to survive the semiclassical limit.
Again, the probability distribution function $P(\Psi)$ and the cumulative
distribution function $I(\Psi)$ are in good agreement with the Gaussian
function as for the scarred states shown before. Thus, for most of the
eigenstates, even though they are scarred, the Shnirelman's theorem
applies in the semiclassical limit. 

The bouncing ball state is a very special feature of the stadium billiard. 
It is well know that due to the existence of a large number of bouncing
ball states, the level spacing statistics in the stadium billiard (for
$\epsilon=1$ or larger $\epsilon$) deviates from the GOE of random matrix
theory at lower energy range \cite{Weid92,SSCL93}. We have calculated the 
energy level statistics
by using the first 2,000 levels for stadium with $\epsilon=1$, the
best-fitting gives rise to the Brody parameter $\beta= 0.83$, which is
comparable with the experimental result ($\beta=0.82$) of Gr\"af {\it et
al}\cite{Weid92}. This number is evidently far from that value of GOE
($\beta=1$) of random matrix theory. Therefore, as the last example of the
high-lying eigenstates, we would like to show a representative of the
bouncing ball states. 

The bouncing ball state shown in Fig. 14 has eigenvalue of $k=$1329.477
057.  As it should be, this energy is very close to the eigenenergy of the
rectangle billiard with the side length of $1$, which has quantum number
$m=13,n=423$, and thus the eigenvalue $k_{mn} = 1329.52112$.  Our
numerical results demonstrate that, almost all the bouncing ball eigenstates'
energy approximately obey this law. At such a high energy level, we have
observed many bouncing ball states, for instance, the three nearly
degenerate consecutive states at $k=1328.1266, 1328.1278$ and $1328.1315$
showing very distinct bouncing ball signature. For these states, the
probability density distribution function deviates strongly from Gaussian. 

Finally, we would like 
point out that although the bouncing ball states survive
the semiclassical limit, the  fraction of the bouncing ball 
states to the total number states will nevertheless vanish in the 
semiclassical limit. (For more detail about the fraction of the bouncing 
ball states, please see the recent two papers by Tanner \cite{Tanner97} 
and B\"acker {\em et al}\cite{BSS97}.) Therefore, the deviation of the 
energy 
level statistics from GOE will eventually disappear in the semiclassical 
limit.

\section{Discussions and conclusions}

In this paper, we have improved Heller's plane wave decomposition method,
with the improved method we are able to calculate the very high-lying
eigenstates, as high as about 1 millionth, of the stadium billiard with a
very high accuracy (better than $10^{-4}$ of the mean level spacing). By
using the approximate semiclassical quantization criterion
Eq.(\ref{action}), we have systematically and extensively searched and 
collected the
scarred states in a very wide range of energies, varies from ground state
to that in the very far semiclassical regime. 

Our numerical results demonstrated that the semiclassical criterion
(\ref{action}) works very well and sometimes even accurate within one mean
level spacing. Furthermore, we have analyzed the scaling property of scar
with $\hbar$. We found that the maximal integrated density fluctuates from
scarred state to state, but the locally averaged intensity scales with
energy in different way for different types of scars. For the
diamond-shape scar, the averaged maximal integrated density does not
depend on $\hbar$, which implies that this type of scar survives the
semiclassical limit. This finding confirms qualitatively and
quantitatively Ronbik's theory of scars\cite{Robnik89}. 

In addition, we have also discovered that some type of scars, e.g. the 
V-shape
and the horizontal bouncing ball scars, their geometrical structures
such as the scar profile and the position of the self-focal point etc. can
be determined well by Bogomolny's theory.  The width of the scar shrinks
approximately with $\hbar^{1/2}$ for individual eigenstate as predicted by
this theory, although the theory is an averaging result of many
consecutive eigenstates. 

Even though the eigenstates in the very high semiclassical limit are
scarred, the probability distribution function is nevertheless an
excellent Gaussian function, which verifies the Shnirelman's theorem. 
 
As illustrated by the examples in this paper, the wavefunctions of
eigenstates contain so rich structures that the nowadays semiclassical
theory cannot predict all of them in detail. It is still a long way to go
for us to be able to predict the wavefunction structures of a given
individual eigenstate. But, we believe that the periodic orbits theory
could contribute more in this direction. 

\bigskip 
\bigskip 
\section*{Acknowledgments}

We would like to thank the referees for helpful and stimulating 
suggestions and comments. B. Li would like to thank Professor Dr. Marko 
Robnik for
discussions. He is also very grateful to Professor Dr. Felix Izrailev for
helpful discussions during the STATPHYS 19 in Xiamen (1995) and during his
visit in Como (1996).  This work was supported in part by the Hong Kong 
Research Grant Council grants and the Hong Kong Baptist University FRG. 
The work done in Slovenia was supported by the Ministry of Science and
Technology of Republic of Slovenia.

\newpage
\section*{Figure captions}
\bigskip
\bigskip

\noindent
{\bf Figure 1a-f:} The probability density plots of the wavefunction for 6 
representative scarred eigenstates of 
odd-odd parity. The wavenumbers $k$ are given in the figure. The highest 
one has $k = 1328.153849$, which corresponds to the
index of 250,034 by using the Weyl formula (odd-odd), which corresponds to 
approximately the 1,001,408th eigenstate for the total billiard. The 
scar is apparently supported by the diamond-shape periodic orbit shown in 
Fig. 3. The stadium has the parameter of circle
radius $R = 1$ and the straight line length 2. In this figure, the unit 
length is about 211 de Broglie wavelength.
\bigskip
\bigskip

\noindent 
{\bf Figure 2:} The probability distribution function $P(\Psi)$
(top) and the cumulative distribution function $I(\Psi)$ (bottom) of
the eigenstate with $k=1328.153849$ shown in Fig.1f, in comparison with the
Gaussian distribution function (the dotted line).  In the bottom one, three 
small boxed regions are
displayed in the corresponding magnified windows.  It is readily to be seen
that, even though the eigenstate is scarred, its probability distribution
function is an excellent 
Gaussian function. 
\bigskip
\bigskip

\noindent
{\bf Figure 3:} The integral region around the periodic orbit that is 
taken in Eq.(6). The width of the tube is 
$D$  measured perpendicular to the periodic orbit.
\bigskip
\bigskip

\noindent 
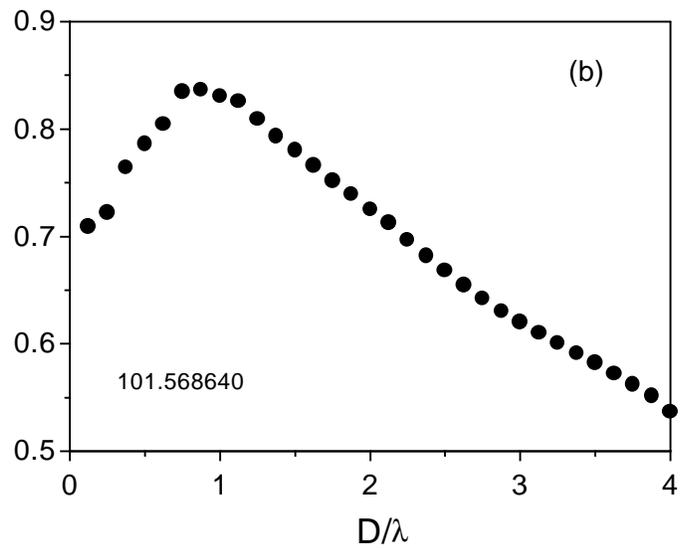
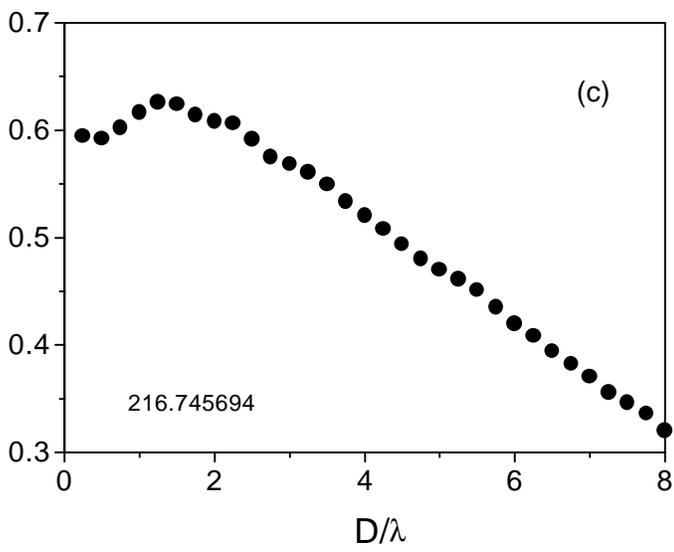
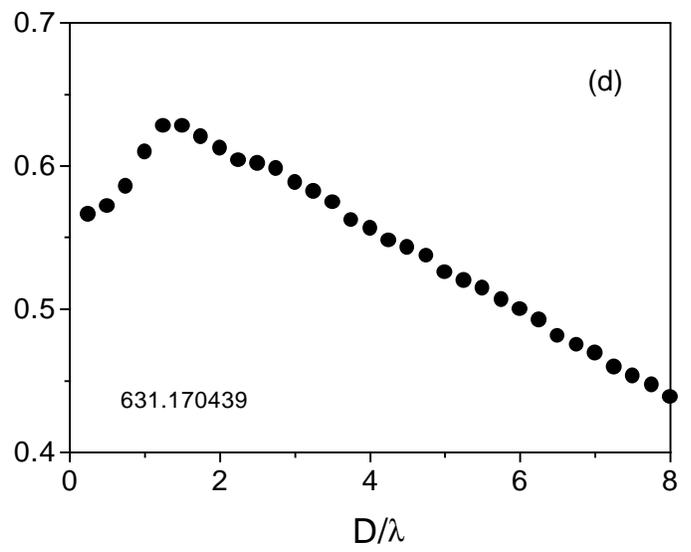
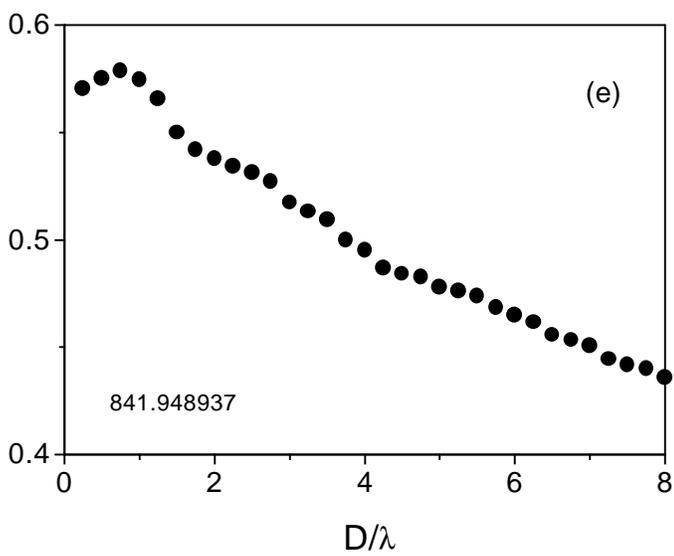
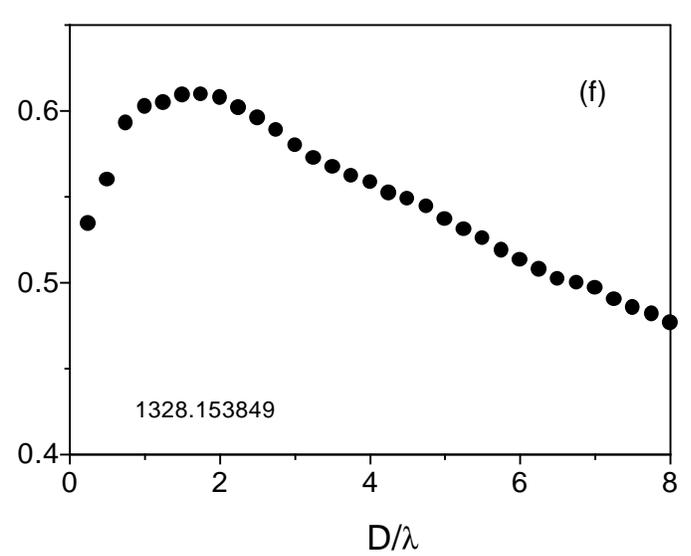
{\bf Figure 4:} The integrated scar intensity profile $I$
versus the width of the integrating tube in unit of the de Broglie
wavelength for the scarred eigenstates given in Figure 1. The wave numbers
are shown at the left bottom of each figure. It is very obvious that
although the eigenvector varies more than 100 times, the maximal
integrated scar intensity does not change too much. In fact, it is
marginally a constant which is about $0.6$.  
\bigskip
\bigskip

\noindent
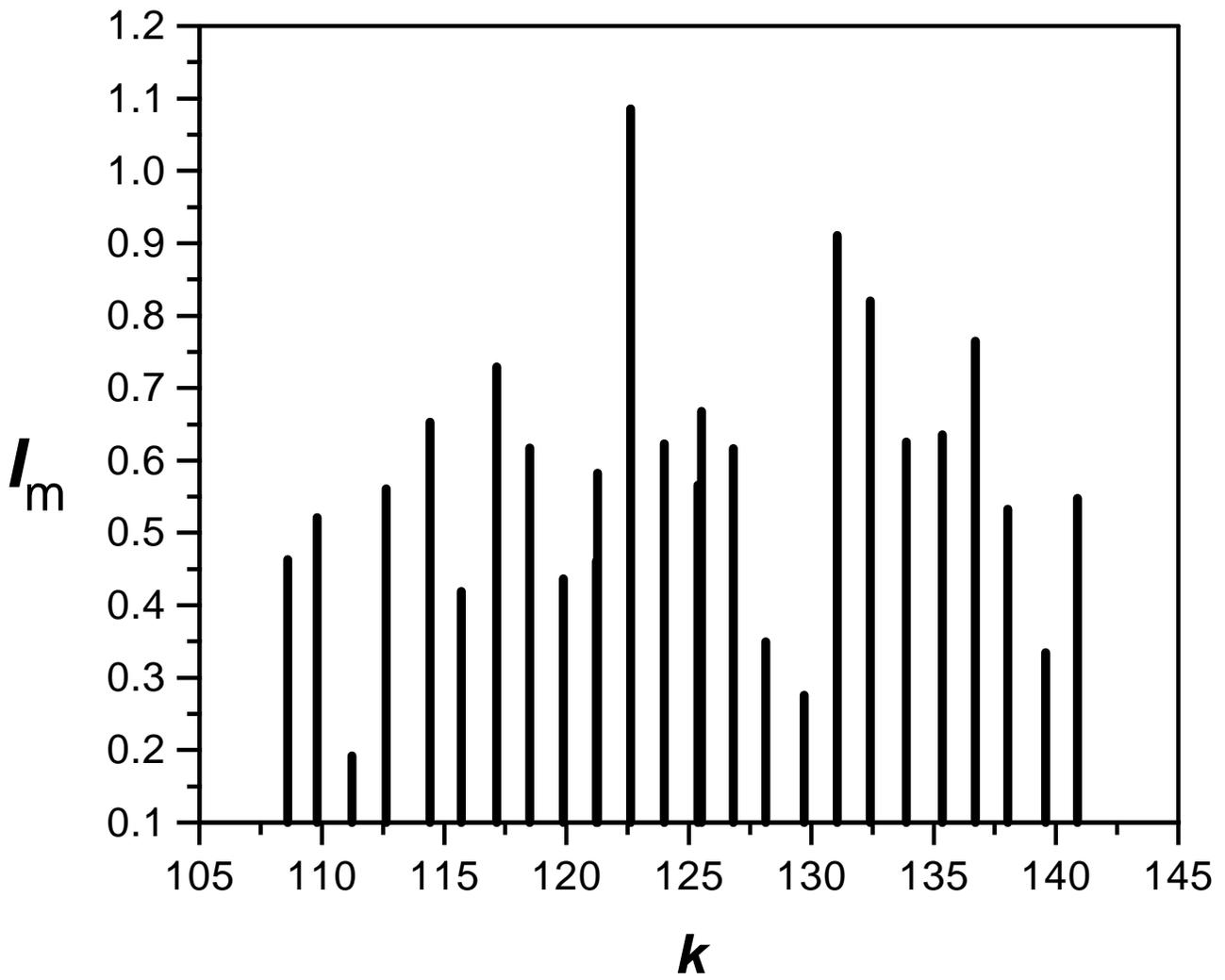
{\bf Figure 5:} The maximum of integrated scar 
intensity versus the  wave number $k$ around $k=125$ for 24 consecutive 
scarred states. The type of 
scar is the same as shown in Fig. 1, i.e. the diamond-shape scar. It should 
be noted that the interval of the wave number between two 
consecutive scarred 
states is very close to $2\pi/{\cal L}$ (= 1.40496), as predicted by the 
semiclassical quantization condition Eq.(\ref{action}). 
\bigskip
\bigskip

\noindent 
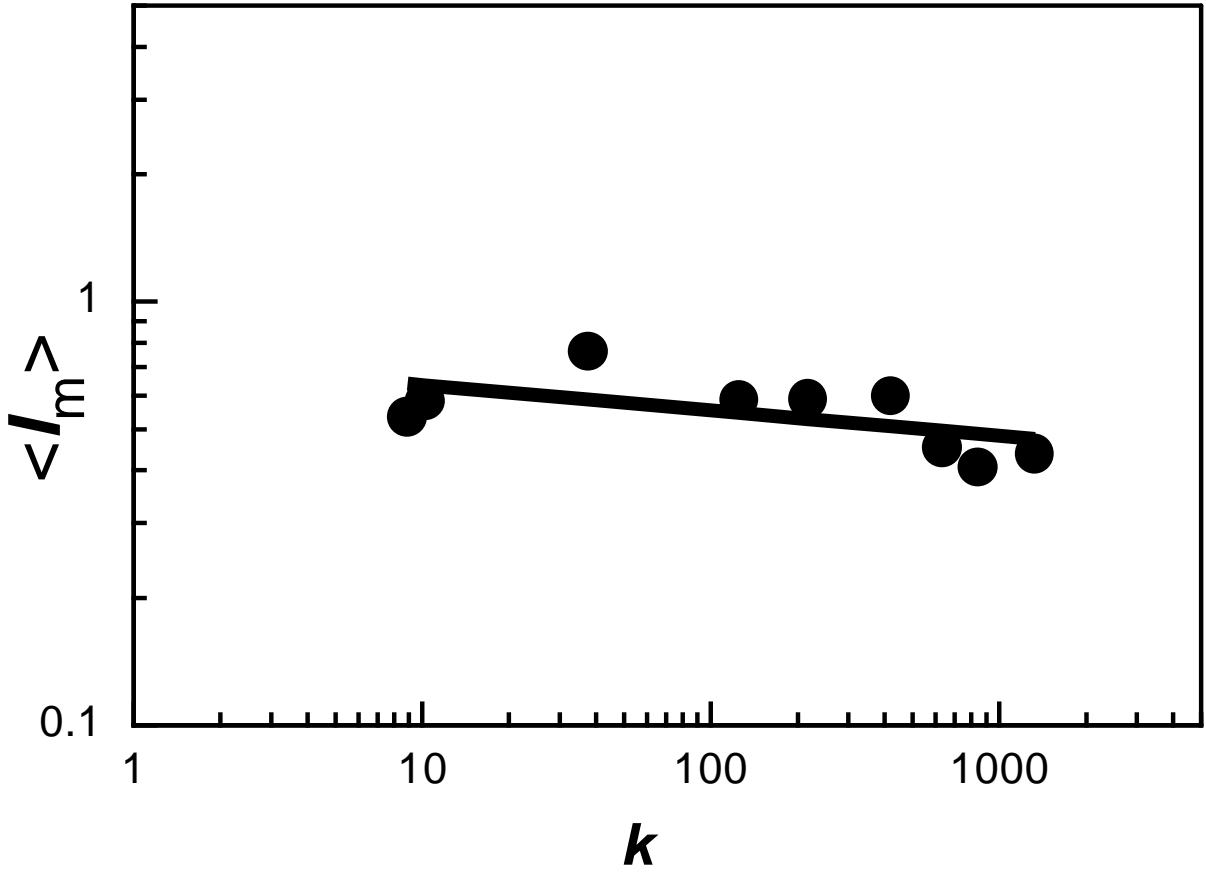
{\bf Figure 6:} The locally averaged (over a small group of
consecutive scarred states) maximum of the integrated scar intensity versus
the wave number $k$. The solid circle represents the numerical data, and the
solid line is the least-square fitting, which is $0.73/k^{\alpha}, 
\alpha
=0.06\pm0.03$. $\alpha$ is very close to zero means that this type of scar
survives the semiclassical limit.  
\bigskip
\bigskip

\noindent 
{\bf Figure 7:} The probability density plot for a scarred
egienstate with wave number $k=$ 1328.093 482, which corresponds to index
250,012 (odd-odd), and to the index about 1,001,317 for the total
billiard. The scar is obviously supported by the V-shape periodic
orbit. There is a clear so-called self-focal point at about 
$x'\approx 1.85$. $x'$ is measured from the center of the straight line 
segment at the billiard boundary. This agrees very well with Bogomolny's 
theoretical prediction (for more please see text).
\bigskip
\bigskip

\noindent
{\bf Figure 8:} The same as Fig. 2 but for the scarred state show in 
Fig. 7.
\bigskip
\bigskip

\noindent
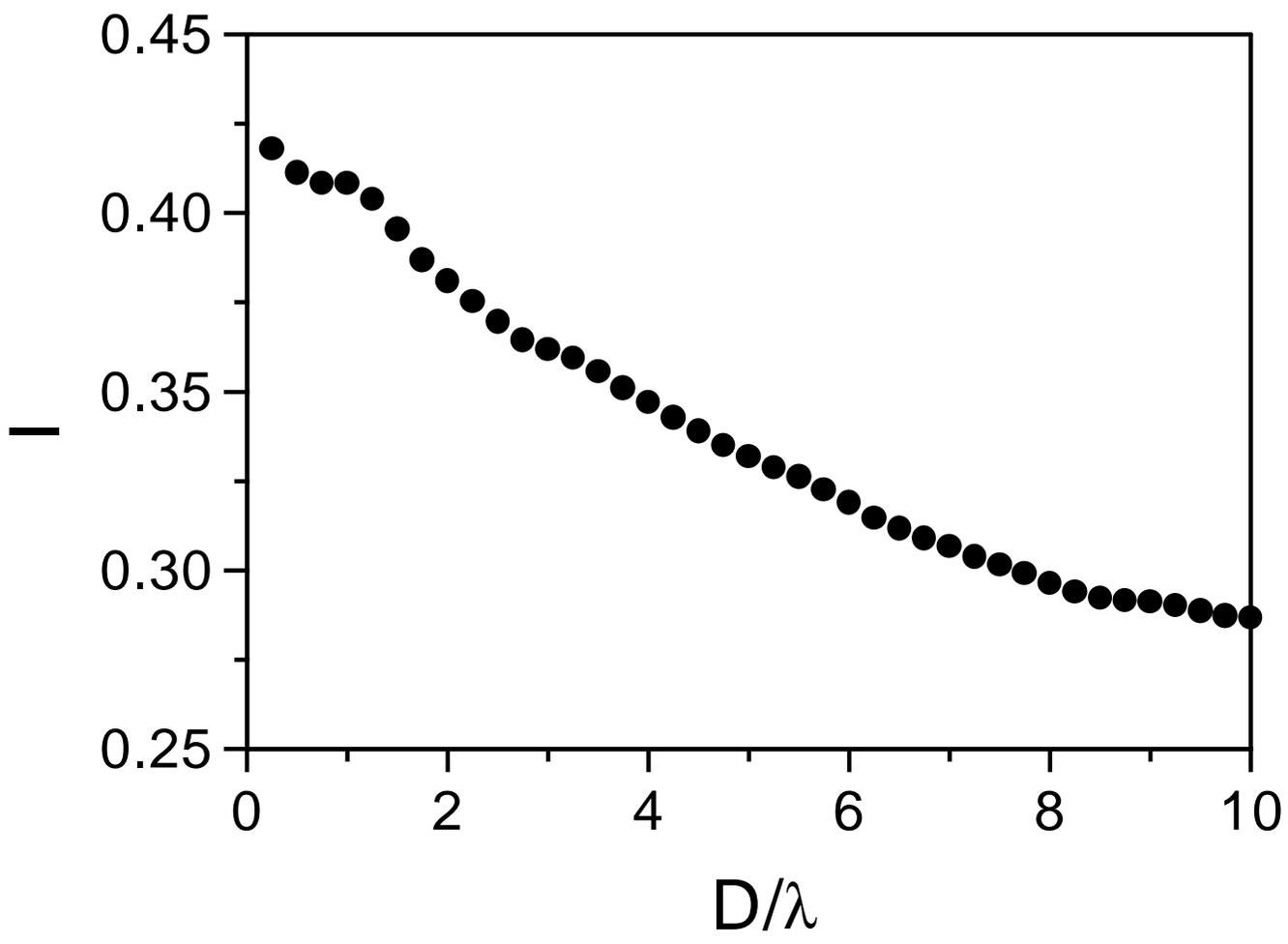
{\bf Figure 9:}
The integrated scar intensity profile $I$ 
versus the width of 
the integrating tube in unit of the de Broglie wavelength for the 
eigenstate drawn in Fig. 7.
\bigskip
\bigskip

\noindent
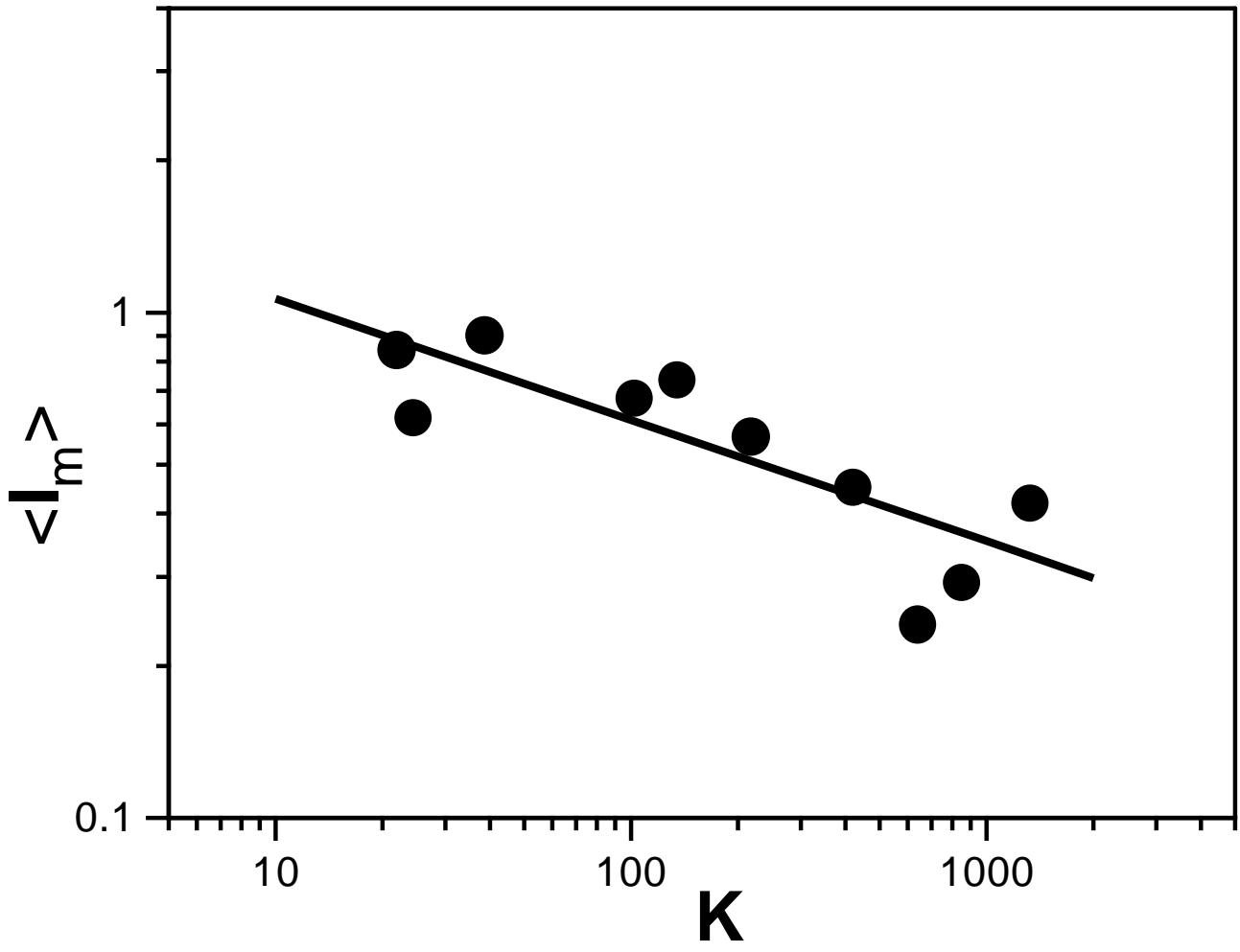
{\bf Figure 10:} 
The locally averaged (over a small group of 
consecutive scarred states) maximum of integrated scar intensity
versus wavenumber $k$. The solid circles represent the numerical data, and 
the 
solid line is the least-square fitting, which is $1.85/k^{\alpha}, 
\alpha =0.24\pm 0.06$. $\alpha$ differs from zero significantly, which
indicates that this type of scar cannot survive the semiclassical 
limit. It 
will vanish eventually if we go to even deeper in the semiclassical regime. 
\bigskip
\bigskip

\noindent
{\bf Figure 11a-h:} 
The probability density plots of wavefunctions for 8 
representative scarred eigenstates 
(odd-odd parity) supporting by the horizontal periodic orbit.
The stadium has the parameter of circle
radius $R = 1$ and the straight line length 0.4.
The wave numbers $k$ are given in the figure. The highest 
one is $k =$ 800.303 338, which corresponds to 
index 49,858 using the Weyl formula (odd-odd), thus it corresponds to 
approximately the 200,445th eigenstate for the total billiard. 
The shape of the pronounced wavefunction around the periodic orbit as 
well as the self-focal point's position can be estimated
approximately by Bogomolny's theory (see text).
\bigskip
\bigskip

\noindent
{\bf Figure 12a-h:} The geometry of the scars calculated from 
Bogomolny's semiclassical theory. The corresponding scarred states' wave 
numbers are presented in the figure. The width of the scar is determined by  
$\frac{C}{\sqrt{k}}\sqrt{|0.24-x^{'2}|}$,  here the constant is 
so chosen that the geometry of the first one ($k=$11.994 542) is 
approximately overlap the scar's geometry shown in figure 11a. 
Accidently, in our calculation $C=(11.994542)^{1/2}$. The goodness of the 
Bogomolny's theory is clearly seen, in particular, at the very high 
eigenstates 
such as that shown in Fig 11g and Fig. 11h. Both the self-focal 
point, which locates at approximately $x' \approx 0.5$, and the scar 
shape are roughly captured by his theory.
\bigskip

\noindent
{\bf Figure 13a-b:} The probability density plots for two very high-lying 
scarred states. The scar are supported by different periodic orbits. The 
wave numbers 
are given in the figure, and the sequential number are about $250,002$ and 
$250,019$ (odd-odd), which correspond to 1,001,280 and 1,001,345, 
respectively, when all parities are taken into account.
\bigskip

\noindent
{\bf Figure 14:} One representative bouncing ball state with $k=$ 
1328.477 057 which corresponds to the sequential number about 250,533 
(odd-odd) and 1,003,405 (total billiard), respectively. Please note that 
the eigenvalue $k$ is very close to the eigenvalue of an $1\times1$ 
rectangle billiard of the quantum number $m=13$ and $n=423$, thus $k_{nm}
=$ 1329.52112. 

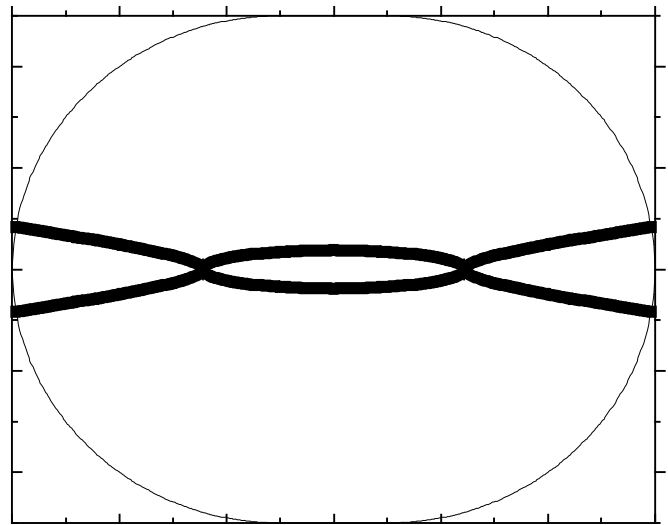
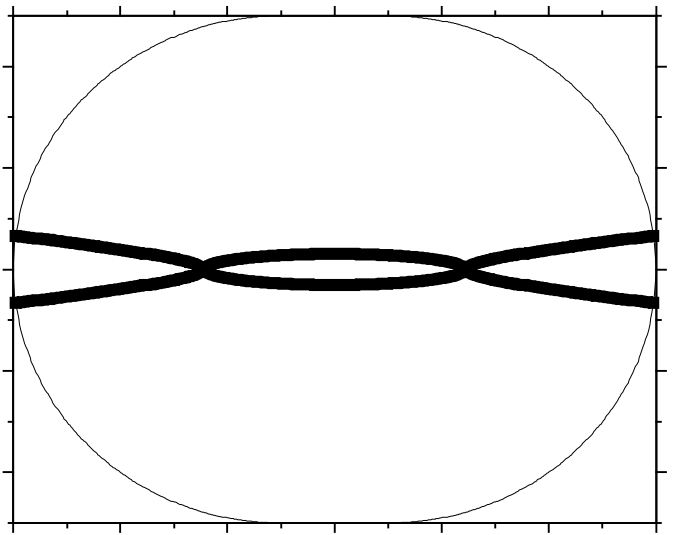
\end{document}